\documentclass[preprint,3p,12pt]{elsarticle}

\usepackage{amssymb}
\usepackage[colorlinks=true]{hyperref}
\biboptions{numbers,sort&compress}
\journal{Nuclear Physics B}

\begin{document}

\begin{frontmatter}

%% Title, authors and addresses

%% use the tnoteref command within \title for footnotes;
%% use the tnotetext command for theassociated footnote;
%% use the fnref command within \author or \address for footnotes;
%% use the fntext command for theassociated footnote;
%% use the corref command within \author for corresponding author footnotes;
%% use the cortext command for theassociated footnote;
%% use the ead command for the email address,
%% and the form \ead[url] for the home page:
%% \title{Title\tnoteref{label1}}
%% \tnotetext[label1]{}
%% \author{Name\corref{cor1}\fnref{label2}}
%% \ead{email address}
%% \ead[url]{home page}
%% \fntext[label2]{}
%% \cortext[cor1]{}
%% \affiliation{organization={},
%% addressline={},
%% city={},
%% postcode={},
%% state={},
%% country={}}
%% \fntext[label3]{}

%\title{The Effect of Sintering Temperature on the Microstructure and Material Properties Molded Martian and Lunar Regolith}

\title{Effect of Sintering Temperature on Microstructure and Mechanical Properties of Molded Martian and Lunar Regolith}

%% use optional labels to link authors explicitly to addresses:
%% \author[label1,label2]{}
%% \affiliation[label1]{organization={},
%% addressline={},
%% city={},
%% postcode={},
%% state={},
%% country={}}
%%
%% \affiliation[label2]{organization={},
%% addressline={},
%% city={},
%% postcode={},
%% state={},
%% country={}}

\author[inst1]{Peter Warren}
\author[inst1]{Nandhini Raju}
\author[inst1]{Hossein Ebrahimi}
\author[inst1]{Milos Krsmanovic}
\author[inst1]{Seetha Raghavan}
\author[inst1]{Jayanta Kapat}
\author[inst1]{Ranajay Ghosh}

\affiliation[inst1]{organization={University of Central Florida, Department of Mechanical and Aerospace Engineering},%Department and Organization
 addressline={4000 Central Florida Blvd}, 
 city={Orlando},
 postcode={32816}, 
 state={Florida},
 country={USA}}

\begin{abstract}
 
Cylindrical specimens of Martian and Lunar regolith simulants were molded using a salt water binder and sintered at various temperatures for comparing microstructure, mechanical properties and shrinkage. Material microstructure are reported using optical microscope and material testing is done using an MTS universal testing machine. The experimental protocol was executed twice, once using Mars global simulant (MGS-1), and once using Lunar mare simulant (LMS-1). The specimens were fabricated via an injection molding method, designed to replicate typical masonary units as well as the green stage of Binder Jet Technique, an important additive manufacturing (AM) technique. Results show that for both the Martian and Lunar regolith that the optimal sintering temperature was somewhere between 1100$^{\circ}$C and 1200$^{\circ}$C. The compressive strength for both the Martian and Lunar masonary samples, that received optimal sintering contditions, was determined to be sufficient for construction of extraterrestrial structures. The work  demonstrates that both the Martian and Lunar regolith show potential to be used as extra terrestrial masonary and as parent material for extra terrestrial BJT additive manufacturing processes. 
\end{abstract}

%%Graphical abstract
%\begin{graphicalabstract}
%\includegraphics[scale=.55]{figs/Intro.pdf}
%\end{graphicalabstract}

%%Research highlights
%\begin{highlights}
%\item Research highlight 1
%\item Research highlight 2
%\end{highlights}

\begin{keyword}
%% keywords here, in the form: keyword \sep keyword
Martian Regolith \sep Lunar Regolith \sep Additive Manufacturing \sep Ceramic additive manufacturing \sep Space colonization \sep ISRU 

%\sep Digital Image Correlation \sep Compressive Strength \sep Material Testing
%% PACS codes here, in the form: \PACS code \sep code
%\PACS 0000 \sep 1111
%% MSC codes here, in the form: \MSC code \sep code
%% or \MSC[2008] code \sep code (2000 is the default)
%\MSC 0000 \sep 1111
\end{keyword}

\end{frontmatter}

%% \linenumbers

%% main text
\section{Introduction}
\label{sec:1}
\begin{figure}[ht]
	\centering
		\includegraphics[scale=.50,trim={0 0.5cm 0 3cm},clip]{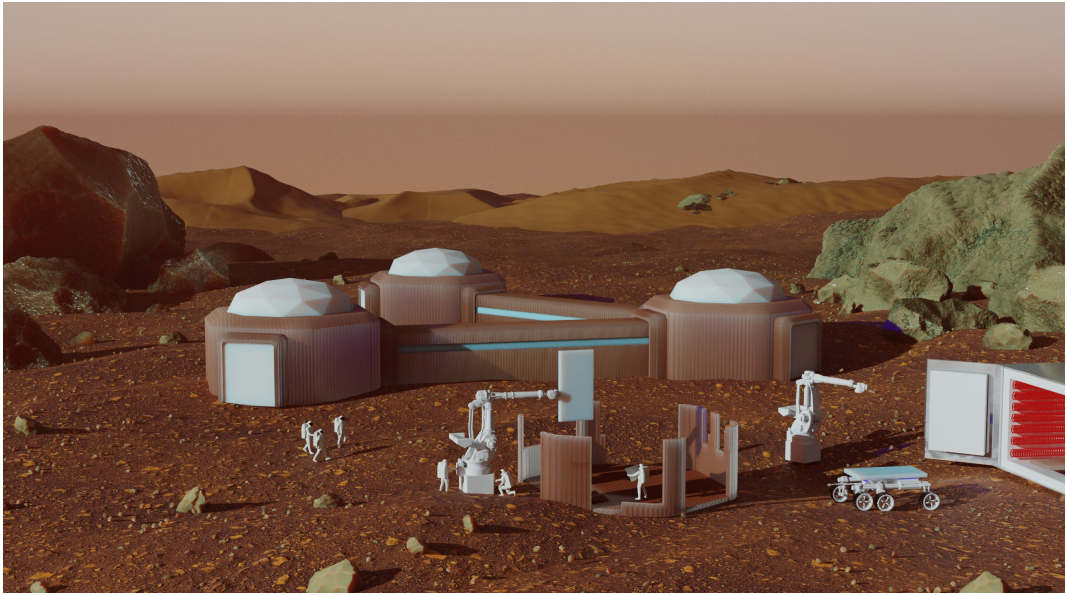}\vspace{-35pt}
	\caption{Depiction of a concept habitat of a Martian outpost. The outpost shows complexity of shapes and materials that must be constructed using ISRU. Additively manufactured sub-components can speed up time and reduce energy for assembly procedure that would involve work by astronautss and robots.}
	\label{Intro}
\end{figure}

The prohibitive cost of transporting raw materials from the Earth to either the Moon or Mars is the driving factor for pursuing in-situ resource utilization (ISRU). The lack of oxygen, extreme thermal cycles, solar and cosmic radiation are significant challenges when attempting to construct a base in these harsh environments \cite{wilhelm2014review, bodiford2005situ, simonsen1990space, ruess2006structural}. Such bases, would most likely be constructed using a variety of materials, architectures and shapes as depicted in Figure \ref{Intro}. Among these, and primarily due to their natural abundance, Martian and Lunar regolith would remain the first choice for parent materials. Regolith has also shown promise in it's ability to withstand the thermal cycles and radiation in their respective environments \cite{miller2009lunar, simonsen1991radiation, slaba2013radiation}. However, using regolith for construction remains a formidable challenge for ISRU. Several materials in the masonry processing typically done on earth are missing in these harsh and desolate extra terrestrial environments in addition to the environmental conditions themselves. 

In spite of this, many techniques have been proposed to utilize this material for realistic construction. These include using a brick making process that implemented a polymer material to bind the regolith powder together \cite{chow2017direct}. Work has been done to implement solar rays to complete the sintering process in a layer by layer fashion \cite{meurisse2018solar}. Binding methods and materials have also been examined by recent research. Some works have been done that introduced a naturally occurring biopolymer (guar gum) to act as a binding agent \cite{dikshit2021space, dikshit2020microbially, kumar2020bacterial}. The biopolymer could be potentially grown in situ, and the bricks constructed using this binding agent showed promising results. Some studies have experimented with bricks formed using liquid sulfur\cite{wan2016novel, khoshnevis2016construction,troemner2020marscrete}. The process is similar to concrete formation, so it is often referred to as "MarsCrete."

These exciting advances suggest the crucial benefit of an organic or inorganic binding material to realize higher quality masonary. More interestingly, such binder materials can also form the basis of extra terrestrial binder jet printing technology (BJT). BJT is an additive manufacturing (AM) method that was developed in the early 1990s and is currently being used to manufacture both ceramic and metallic components \cite{li2020metal, raju2021material}. It is now steadily gaining attention due to its versatility. BJT has a less complicated design that does not use any high powered lasers, high printing resolution, and has demonstrated low cost scalability as a manufacturing platform \cite{ziaee2019binder}. The components are manufactured by accurately distributing a binder material onto a powder bed, lowering the powder bed, rolling out a new layer of powder, and then distributing the binder material again \cite{gibson2021binder, ziaee2019binder}. BJT has been implemented using regolith as a powder bed and salt (NaCl) water as a binder in order to additively manufacture components \cite{ceccanti20103d, cesaretti2014building}. Although no post print sintering was completed, the work still showed the capability to produce green components from lunar regolith and the binder material. The green part or green state refers the point in the manufacturing process at which the part has been printed but not sintered. This means the particles are only held together by the binder material, so the part is still fragile. After the initial print, the green component must undergo a sintering phase in order to bond the powder material together by reaching near melting temperatures of the powder material. 

Promising as this technology is, binder material and sintering play a crucial role in BJT's success. More interestingly, other binder based technologies mentioned above can also potentially benefit from sintering since significant densification, which is critical towards high final strength occurs during the sintering process \cite{karl2020sintering}. However, the sintering also causes significant microstructural changes and sample shrinkage. Thus, it has a critical impact on the overall accuracy and quality of the final manufactured product.

Considering this centrality of sintering, this work investigates the effect of sintering temperature on green regolith parts made using  a simple yet potentially  sustainable salt water binder. The microstructural evolution of martian and lunar regolith samples with sintering temperature, shrinkage and strength improvements are reported in detail.

\section{Materials and Methods}
\label{sec:2}

\subsection{Regolith Simulant}

There have been about 400 kg of Lunar regolith retrieved and brought to earth during the apollo missions \cite{mclemore2009need}. There has not currently been any Martian regolith retrieved from mars, all study has been performed in situ. Therefore the quantity of Lunar regolith is insufficient for significant research and the quantity of Martian regolith is nonexistent \cite{mclemore2009need}. This has driven the need for simulants. There have been around 30 different variants of the Lunar regolith and 40 types of Martian regolith developed \cite{altun2021additive, liu2021situ}. Variations in the regolith could cause variations in the final material properties of any manufactured component. For this work, we will be using MGS-1 and LMS-1 from Exolith labs \cite{cannon2019mars, landsman2020simulated}. Further work to create more realistic and uniform simulants is on going.
\subsection{ISRU Binder}
In this work, a pure distilled H$_{2}$O mixed with a small amount of NaCl is used as the binder material. A water and salt solution has been used before as a binder material for Lunar material \cite{ceccanti20103d, cesaretti2014building}. Salts have been found preserved in the rocks and soil on mars, from when the planet was wetter \cite {clark1981salts}. Sodium is less prevalent on the moon, but trace amounts have been detected in the regolith \cite{mc2001isru}. Water deposits have recently been reported on both mars and the moon, although accessing the water would be challenging \cite{rickman2019water}. The salt-water binder examined in this work is a simplistic and effective ISRU material.

\subsection{Manufacturing method}
The manufacturing method in this work will be injection molding. The molding itself was made from 3D printed PLA components, which are cylindrical in shape. The regolith is mixed with the salt water binder prior to being placed into the PLA printed molds. The regolith mixture is not compressed into the mold, just packed gently into it. Injection molding is a good replication of the binder jet printing method. It is a good replication because the green part is composed of binder and powder for both binder jetting and injection molding.

\subsection{Molding Method}\label{mold}
\begin{figure}[ht]
	\centering
		\includegraphics[scale=.55,trim={2.5cm 3.5cm 2.5cm 3.5cm},clip]{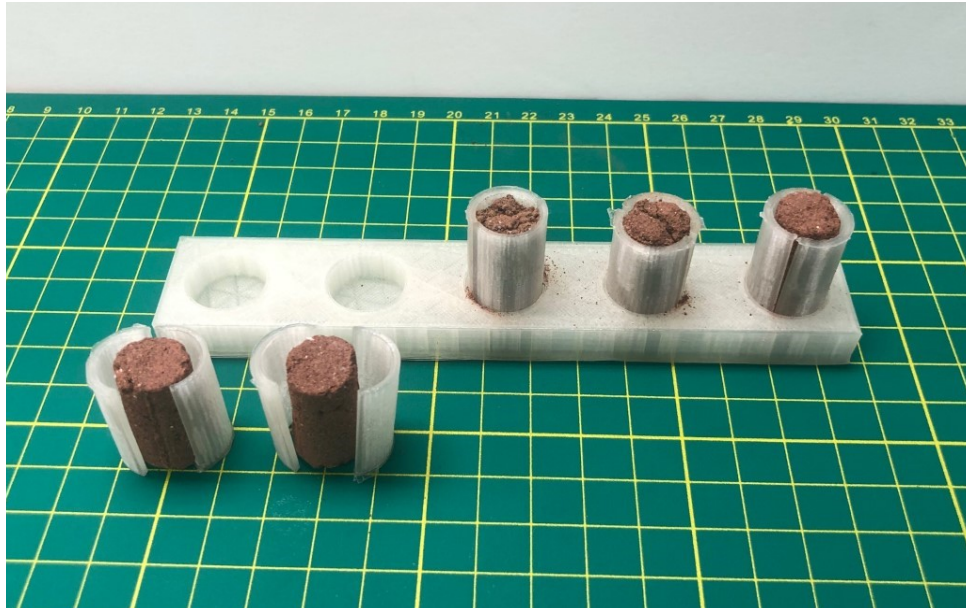}
	\caption{Photograph of the cylindrical samples in the 3D-printed half shell PLA molds. The halved cylindrical style molding method provides easier extraction of the regolith samples. The samples are composed of MGS-1 martian regolith and the binder material is salt water solution.}
	\label{Manu}
\end{figure}

The samples were molded using 3D-printed molds out of a thermoplastic. The molds were 3D-printed on an Ultimaker S5 \textregistered. A Tough PLA material of the translucent color was used for the molds. The samples made were cylindrical in shape and had a length of 1 inch and a diameter of 0.5 inch. The PLA 3D-printed molds can be seen in Figure \ref{Manu}. They have two components one is the cylindrical molds which were 3D-printed in two halves. The other is the base plate which can hold the two halves of the cylindrical mold in a secure fasion. The regolith and binder mixture can then be added to create the desired cylindrical shape.

The binder material was made from pure distilled water and pure iodized salt. The ratio of NaCl to H$_{2}$O was  1/16 or 6.25\% NaCl. The addition of salt to the water made a noticeable difference in the binding of the material together. An attempt was made to produce these samples with just water, and it failed. During extraction from the PLA molds, the regolith cylinders would crumble apart. The addition of salt to the water made the extraction of the samples from the molds much easier and held the regolith together for both the Martian and Lunar versions quite well. 

The regolith and binder ratio was 3.5 parts regolith to 1 part binder. This was selected from observation of the stability of the green part. If too much binder liquid is added the sample will collapse. Also if there is too little binder liquid added the sample will be too brittle and crumble. The regolith and binder mixture was loosely packed into the molds to mimic the behavior of the BJT printing method. They did not sit to allow the water to evaporate either. Immediately after packing the regolith and binder mixtures into the molds they were then extracted and the sintering process was initiated. The same conditions were applied to both the Lunar and Martian regoliths in order to create the data that could be compared upon completion of the manufacturing process.

\subsection{Sintering Protocol}

\begin{figure}[ht]
	\centering
		\includegraphics[scale=.7]{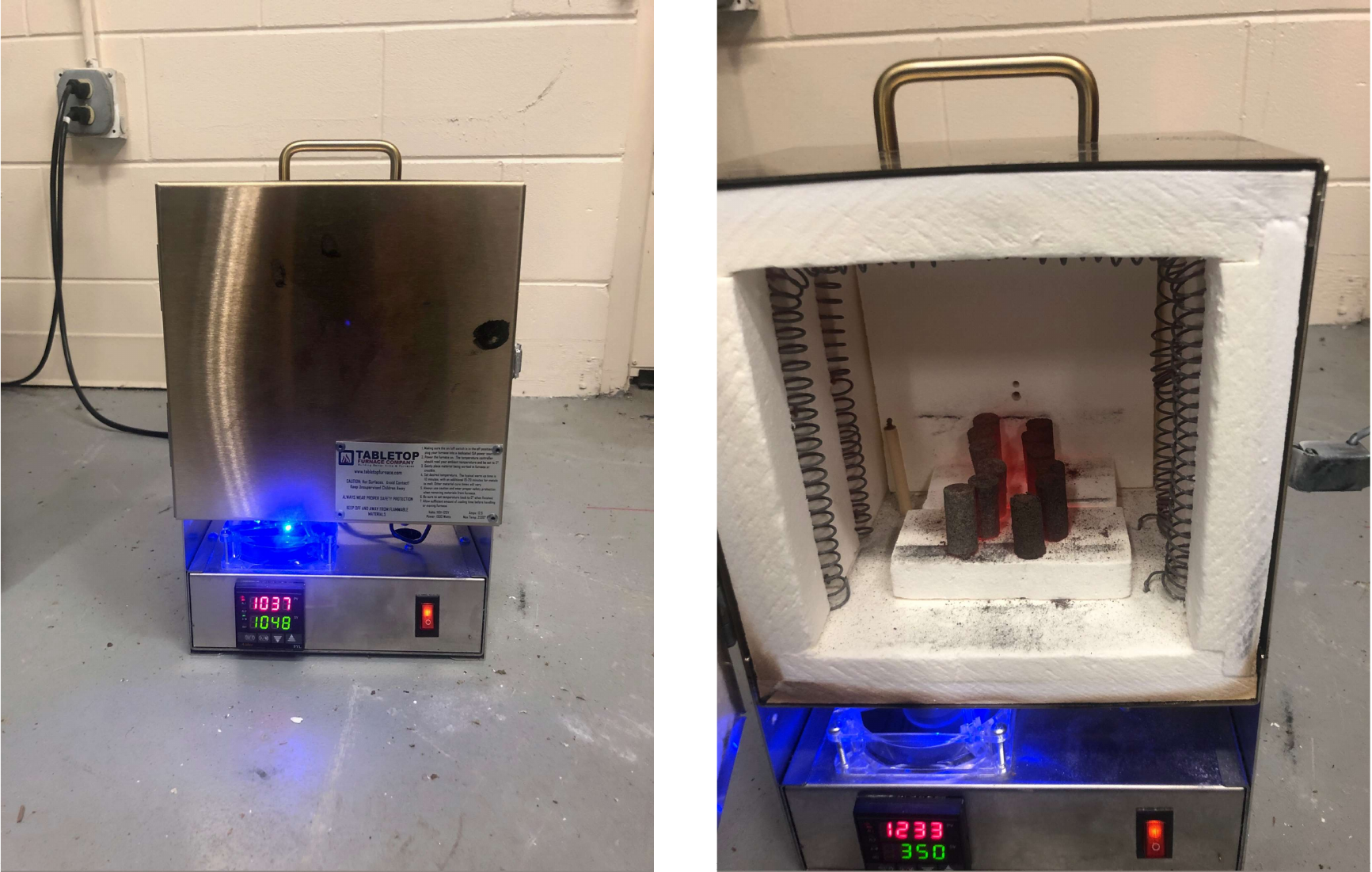}
	\caption{Photograph of the sintering oven used for preparing the samples (Rapidfire Standard Pro I \textregistered). The oven is heated via two electical heating coils that run along the sides and top of the sintering oven. (left) Sintering oven in operation, closed and latched shut. (right) Opened after sintering a set of samples placed on Alumina pads.}
	\label{Furn}
\end{figure}

 The sintering oven used to sinter the samples, is manufactured by the Tabletop Furnace Company and the model is the Rapidfire Standard Pro I \textregistered. The furnace has a maximum temperature of 1200$^{\circ}$C. The oven has 0.2$\%$ accuracy over the entire input range of the furnace. This furnace has two induction coils that run the full length of the sides and top of the furnace. This can be seen in Figure \ref{Furn}. The samples were placed on alumina blocks during the sintering procedures. All of the samples were pre sintered directly after the molding process. The presintering was performed at 200$^{\circ}$C for a total of 1 hour. After completion of the presintering step, the samples were then sintered at high temperatures. The Martian and Lunar sintering conditions were again kept consistent with each other. A total of 15 samples were molded using the aforementioned molding methods. The 15 samples were split into groups of 5 after the presintering was completed. 5 samples were sintered at 1000$^{\circ}$C for 1 hour, 5 samples were sintered at 1100$^{\circ}$C for 1 hour, and 5 samples were sintered at 1200$^{\circ}$C for 1 hour. Again this conducted for both Martian and Lunar regolith samples, creating a total of 30 samples. The heating ramp rate for both the presintering and sintering was 25$^{\circ}$C per minute. The cool down was just completed by turning the oven off and allowing the temperature of the surrounding environment to bring the temperature of the oven back down. The cool down was relatively rapid and took about 10 minutes.

\subsection{Microstructure Inspection}

At every stage of the manufacturing process the material was examined under a microscope. The examinations took place at the powder stage, after the molding phase, and after the sintering phase. The microscope used was a Keyence VHX-900F \textregistered. Images were taken at 10X and 200X magnification. The 10X magnification gives a view of the entire top of the sample, and the 200X magnification provides a  view of the particles that the samples are comprised of. The images will be given in the results section.

\subsection{Material Testing}

\begin{figure}[ht]
	\centering
		\includegraphics[scale=.5,trim={1.5cm 3cm 4cm 1cm},clip]{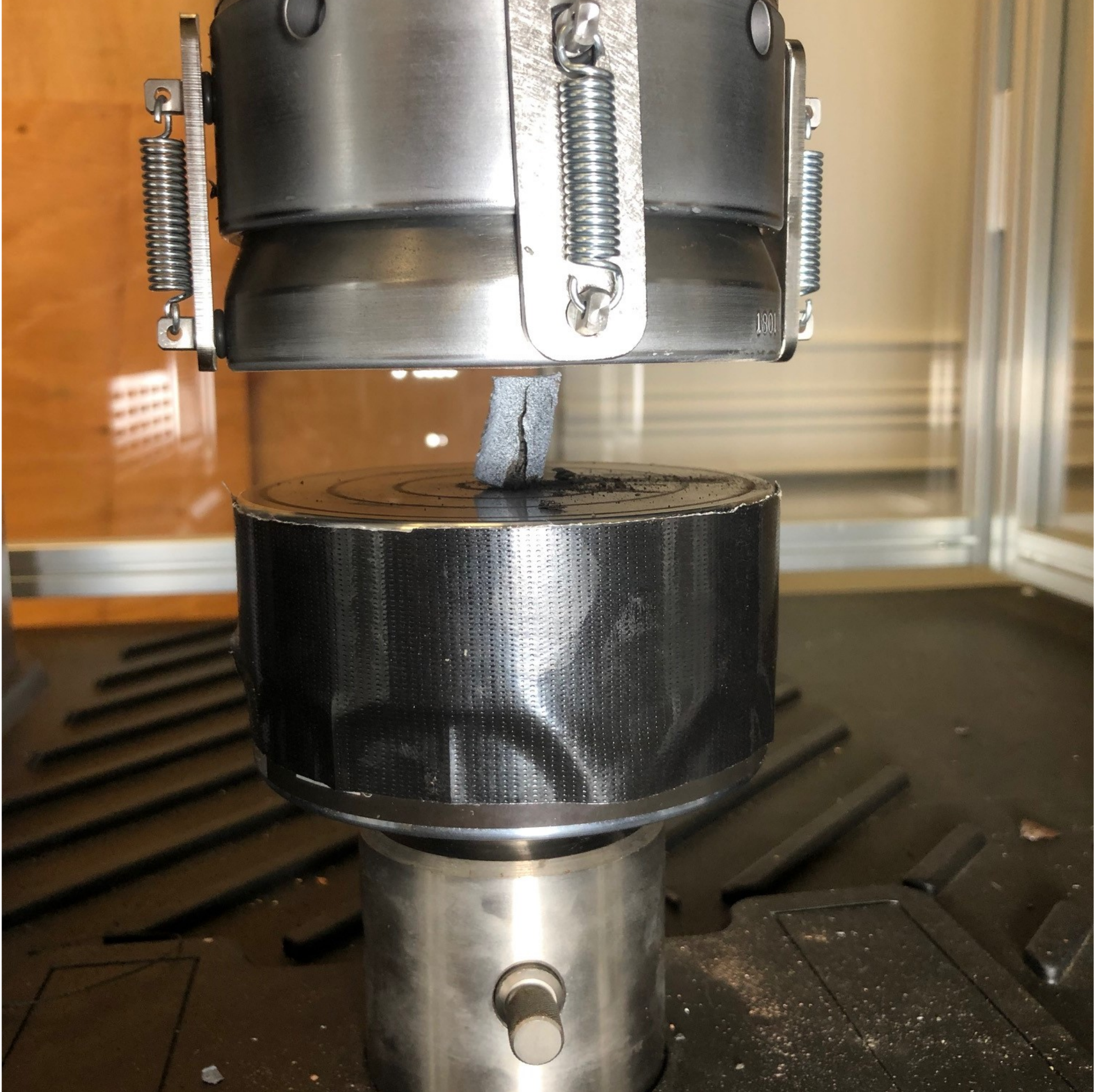}
	\caption{Experimental set-up for the compression testing for the sintered Martian and Lunar Regolith samples. The sample was painted white and then black dots were applied to make it possible to apply DIC measurements. The image was taken after failure of the sample had occurred. There is black duct tape on the lower compression pad to help prevent glare, which effects the DIC system.}
	\label{MTS}
\end{figure}

The material was tested on an MTS machine following the ASTM C1314 - Standard Test Method for Compressive Strength of Masonry Prisms \cite{ASTMC1314-21}. The tensile test machine was an MTS Criterion Model 43, and the load cell was a 50 kN load cell model LPS504 C. Each of the samples were tested giving a sample weight of 5 to each category. The compression modulus was calculated for the samples sintered at 1200$^{\circ}$C, and will be given in the results section. The samples that were sintered at 1000$^{\circ}$C and 1100$^{\circ}$C were too brittle to have a compression modulus of any significance. The material would fail under a relatively small load. The compression modulus testing set up is shown in Figure \ref{MTS}. This image was captured after the compression was performed to illustrate how the sample failed. 

Digital Image Correlation (DIC) was only performed on the samples sintered at 1200$^{\circ}$C. The samples were painted white with black dots on it for Digital Image Correlation (DIC). This is often referred to as the speckle pattern. The DIC system was Dantec Dynamics Q-400 Digital 3D Image Correlation System. The DIC measurements provide a full 3D view of the strain map of the samples while they were under applied compressive load. The tape on the compression pad seen in Figure \ref{MTS} was put there to prevent the glare from affecting the DIC cameras. The compression pads are metal and the lights used for DIC are very bright and the glare affect the data gathered from the camera.

\section{Results and Discussion}

\subsection{Microstructure Images}

\begin{figure}[ht]
	\centering
		\includegraphics[scale=0.65]{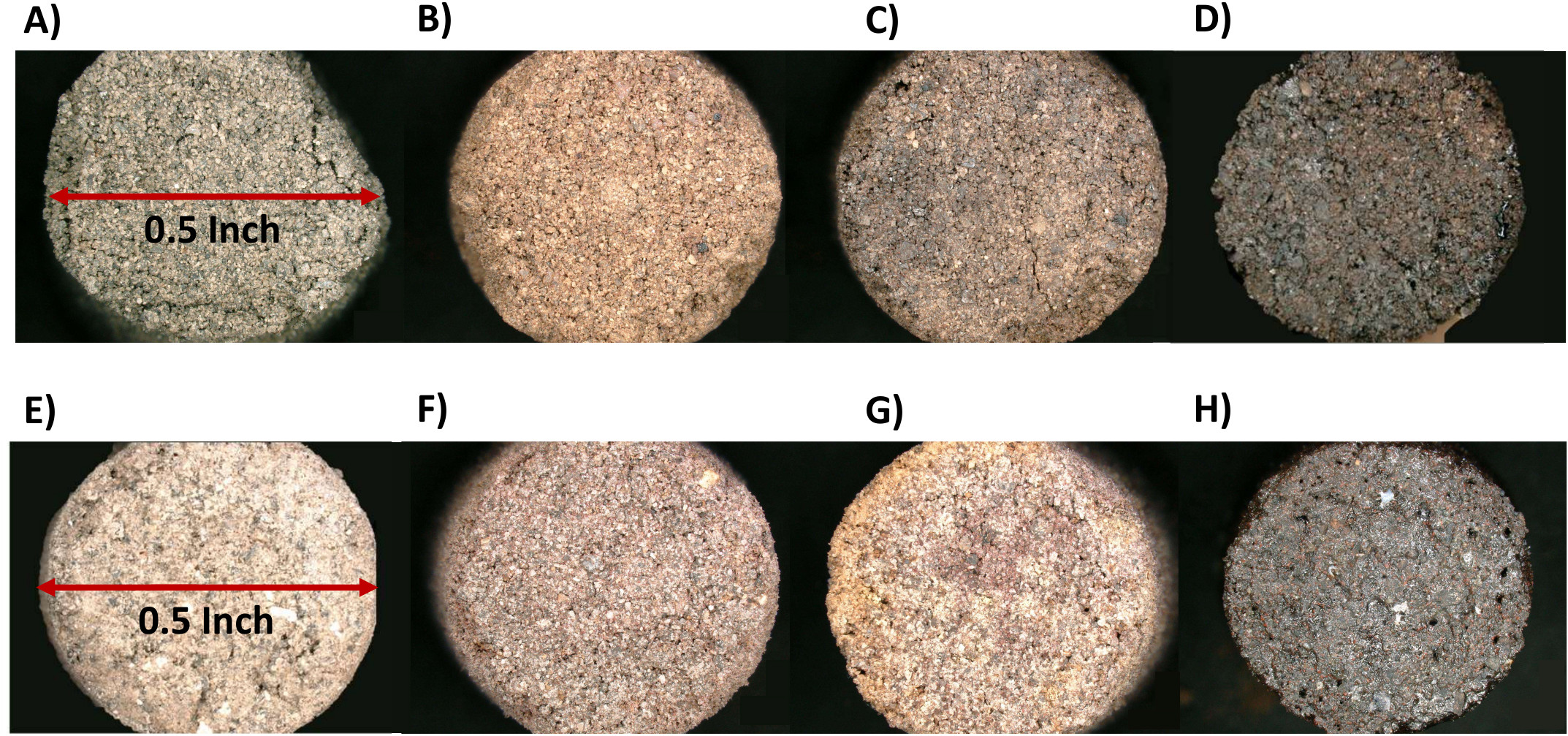}
	\caption{Microscopic images of the top view of the cylindrical regolith brick samples at 10X magnification: (A) Lunar prior to any sintering (green). (B) Lunar sintered for 1 hour at 1000$^{\circ}$C. (C) Lunar sintered for 1 hour at 1100$^{\circ}$C. (D) Lunar sintered for 1 hour at 1200$^{\circ}$C. (E) Martian prior to any sintering (green). (F) Martian sintered for 1 hour at 1000$^{\circ}$C. (G) Martian sintered for 1 hour at 1100$^{\circ}$C. (H) Martian sintered for 1 hour at 1200$^{\circ}$C.}
	\label{Top}
\end{figure}

  Digital images were gathered at both 10X magnification and 200X magnification using the  Keyence VHX-900F \textregistered. In Figure \ref{Top}, we can see the top view of all of the samples. These images were recorded at 10X magnification. In this image, we can see both the Lunar and Martian samples at the green state, sintered at 1000$^{\circ}$C, 1100$^{\circ}$C, and 1200$^{\circ}$C. The images show a very clear distinction between sintering at 1200$^{\circ}$C and the other two lower temperatures for both the Martian and Lunar samples. The difference in coloring between the 1000$^{\circ}$C and 1100$^{\circ}$C is not substantial.

\begin{figure}[ht]
	\centering
		\includegraphics[scale=0.85]{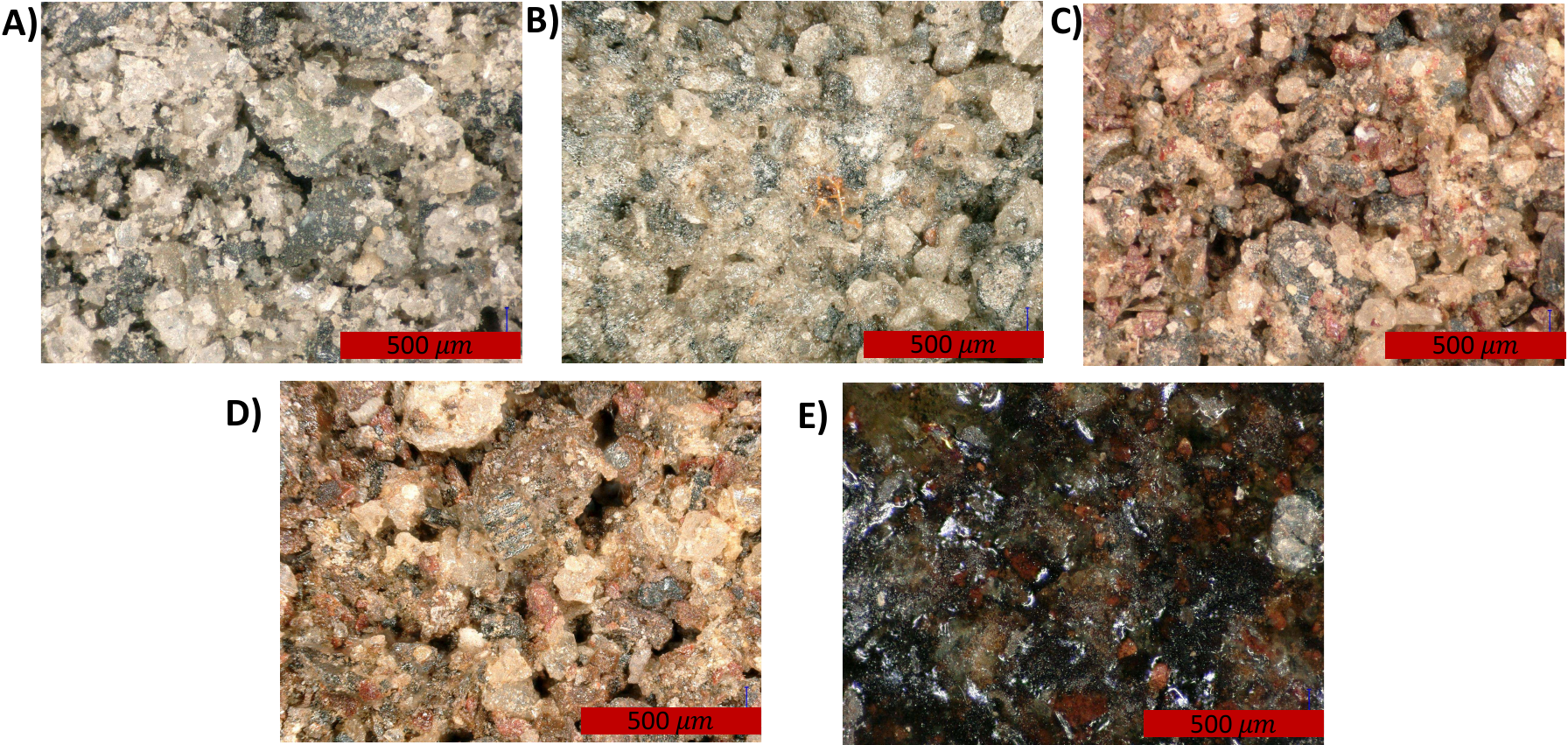}
	\caption{Microscopic images of the Lunar samples collected at 200X magnification: (A) Lunar sample during the green state, after molding, prior to sintering. (B) Lunar sample sintered for 1 hour at 1000$^{\circ}$C. (C) Lunar sample sintered for 1 hour at 1100$^{\circ}$C. (D) Lunar powder prior to any sintering or molding. (E) Lunar sample sintered for 1 hour at 1200$^{\circ}$C.}
	\label{Lunar}
\end{figure}

\begin{figure}[ht]
	\centering
		\includegraphics[scale=0.85]{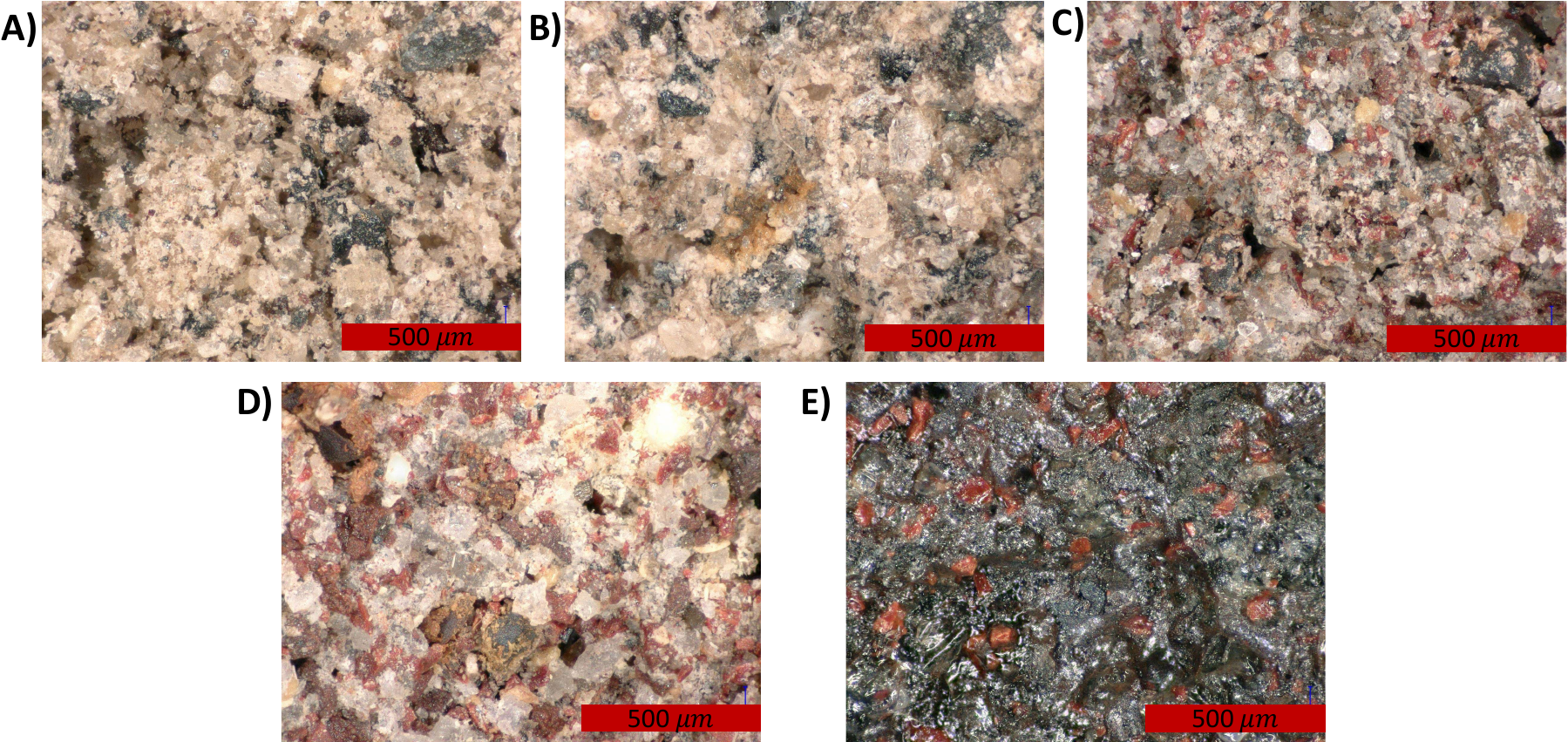}
	\caption{Microscopic images of the Martian samples collected at 200X magnification: (A) Martian sample during the green state, after molding, prior to sintering. (B) Martian sample sintered for 1 hour at 1000$^{\circ}$C. (C) Martian sample sintered for 1 hour at 1100$^{\circ}$C. (D) Martian powder prior to any sintering or molding. (E) Martian sample sintered for 1 hour at 1200$^{\circ}$C.}
	\label{Martian}
\end{figure}

The images that were taken at 200X magnification display the rocky mixture of the composition very well. The images of the Lunar samples at 200X magnification are shown in Figure \ref{Lunar}. In Figure \ref{Lunar}, the powder state, the green state, and the three different sintering conditions are all displayed. These states are shown in the same order for the Martian samples in Figure \ref{Martian}. The temperature sintering conditions applied for one hour of at either 1000$^{\circ}$C, 1100$^{\circ}$C, and 1200$^{\circ}$C. This was completed directly after the presintering portion of the cycle. Again we clearly see the distinction of sintering samples at the temperature of 1200$^{\circ}$C. The 1200$^{\circ}$C sintered image shows a glassy encasement of rocky particles. These samples were also smooth and glass like to the touch. This was true for both the Martian and Lunar samples. This smooth glass like touch of final product has been discussed further in Section \ref{Composition} Mineral Composition.

The comparison between the powder state and the green state shows how the binder material is able to hold the Lunar regolith together. The salt water binder and smaller regolith particles combine to behave like a bonding agent and hold the larger regolith particles together.

In Figures \ref{Lunar} and \ref{Martian} we can also observe the comparison between Lunar and Martian (respectively) regolith samples in the powder form (A) and in the green state (B). The green state is the state at which the binder material was mixed with the powder, but no sintering has occurred yet. The binder material used was a 6.25\% salt-water solution, as explained in Section \ref{mold}. The salt water binder and smaller regolith particles combine to behave like a bonding agent and hold the larger regolith particles together. A more significant green state bond could be formed if the powder material was composed of smaller and more uniform particles. This could be achieved by using a milling machine on the powder prior to addition of the binder material. 

\subsection{Compression Testing}

Compression testing was performed on all of the sintered samples. The 1000$^{\circ}$C and 1100$^{\circ}$C samples did not sinter together well enough to report any significant compressive strength with adequate accuracy. The samples were still quite brittle and would crumble apart if not handled carefully. An attempt was made using a 10 kN load cell on the samples sintered at 1000$^{\circ}$C and 1100$^{\circ}$C, but they would crumble as soon as any significant load was initiated and the data was not useful. The samples sintered at 1200$^{\circ}$C were sturdy enough to report data on the compression modulus and the compressive strength. This data is available for both the Lunar and Martian samples in Table \ref{tab:MTS}. Sintering has been performed on lunar regolith at 1100$^{\circ}$C with successful results, but the hold time was 4 hours \cite{taylor2018sintering}. This same work attempted to sinter at 1050$^{\circ}$C and the results were too brittle to test \cite{taylor2018sintering}.

\begin{table}[ht]
\centering
\caption{Material properties of sintered Lunar and Martian samples.}
\label{tab:MTS}
\begin{tabular}{lccccc} \hline
 & Martian Sample & Lunar Sample \\ \hline
1000$^{\circ}$C Radial Shrinkage Percentage & 1.06$\%$ & 0.84$\%$ \\ \hline
1000$^{\circ}$C Height Shrinkage Percentage & 0.57$\%$ & 1.16$\%$ \\ \hline
1100$^{\circ}$C Radial Shrinkage Percentage & 2.34$\%$ & 1.85$\%$ \\ \hline
1100$^{\circ}$C Height Shrinkage Percentage & 2.52$\%$ & 3.15$\%$ \\ \hline
1200$^{\circ}$C Radial Shrinkage Percentage & 10.54$\%$ & 6.17$\%$ \\ \hline
1200$^{\circ}$C Height Shrinkage Percentage & 16.03$\%$ & 18.13$\%$ \\ \hline
1200$^{\circ}$C Compression Modulus (MPa) & 67.80 & 55.73 \\ \hline
1200$^{\circ}$C Compressive Strength (MPa) & 25.46 & 21.73 \\ \hline
1200$^{\circ}$C Compression Modulus (Ksi) & 9.83 & 8.08 \\ \hline
1200$^{\circ}$C Compressive Strength (Ksi) & 3.70 & 3.15 \\ \hline
\end{tabular}
\end{table}

The Martian samples had a slightly higher compression modulus and compressive strength than the Lunar samples. These values (for both the Martian and the Lunar samples) are in a similar range to that of brick, if not slightly higher. Brick has an compressive strength of around 6 MPa - 82 MPa \cite{gumaste2007strength, page1981biaxial, singh2017bond}. This would indicate that this material is a suitable material for construction of load bearing structures.

\begin{figure}[ht]
	\centering
		\includegraphics[scale=0.9]{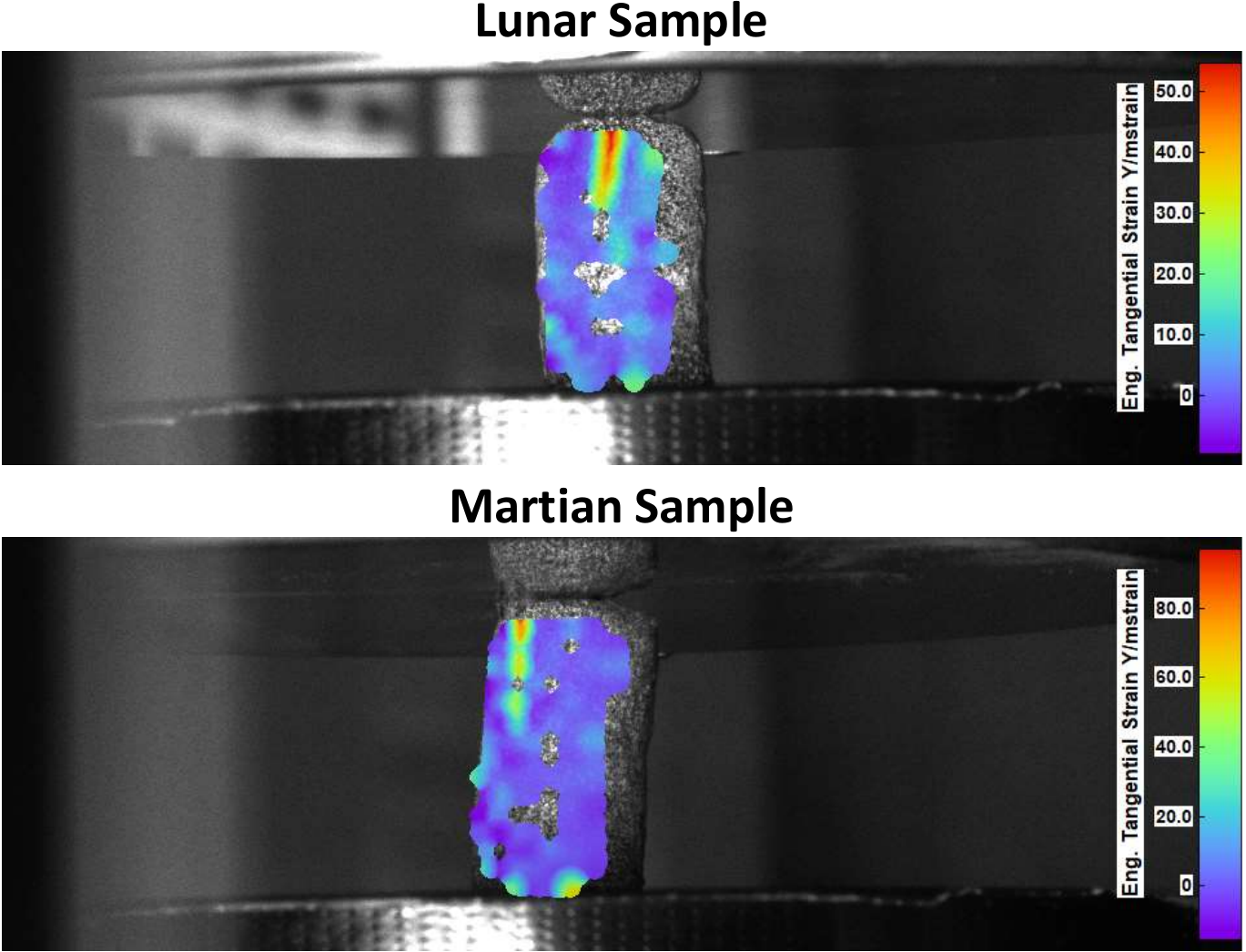}
	\caption{DIC (Digital Image Correlation) of the Martian and Lunar samples. The images were taken just prior to fracture and indicate the early stages of the formation of vertical cracks that cause the failure. The engineering tangential strain is shown, which indicates the ratio of relative displacement to its original distance.}
	\label{DIC}
\end{figure}

The failure mechanism of the samples (both Lunar and Martian) was the formation and rapid propogation of a vertical crack. In Figure \ref{DIC}, we can see DIC (Digital Image Correlation) images. These DIC images provide a view of the strain patterns in real time. The images were taken just before the samples failed. The regions of high strain are in a straight vertical line, beginning at the top of the samples. These high strain regions caused the crack to initiate and propagate down the sample very rapidly. This is an indication that the material is a brittle material and underwent a standard brittle material failure \cite{lemaitre2006engineering}. The crack did not form in a 45 degree line, instead it formed in a 90 degree vertical line. This could be due to the shape of the specimen. During sintering the top of the cylinder was slightly rounded out along the edges. This could have also been caused by stronger bonds forming during sintering in the vertical direction versus the radial direction, leading to an anisotropic material formation. The vertical bonding could be stronger due to the gravitional force acting upon the object during sintering. It was also recorded that there was more shrinkage in the vertical direction during sintering.

The DIC system used was a two camera setup with Dantec Dynamics software. The facet size was 19 pixels, the accuracy was 0.3 pixels, the residuum was 35 gray value, the 3D residuum was 0.4 pixels, and the grid spacing was set to 15 pixels. A glare caused by the shiny compression pads of the MTS was eliminated using black duct tape. The paint on the samples was just a light coat of white spray paint, followed by a very delicate mist of black spray paint. This supplied the necessary speckle pattern for the DIC system to work correctly.

As mentioned in the introduction, other researchers have looked into the mechanical properties of in situ extraterrestrial constructed samples. One study analyzed the flexural strength of Martian material constructed directly from soil compaction \cite{chow2017direct}. A lot of work has been done into the feasibility and ultilization of the Martian concrete formed from liquid sulfur \cite{wan2016novel, khoshnevis2016construction, troemner2020marscrete}. Studies have also been conducted on Lunar regolith based samples as well. One study looked into the effect of powder composition on the compressive strength, where the samples were sintered at 1125$^{\circ}$C \cite{meurisse2017influence}. A solar sintering of Lunar bricks has also been researched, where a polymer was also introduced to the process \cite{meurisse2018solar}. Another work used a naturally occurring biopolymer (guar gum), along with glass fibers to create lunar bricks \cite{dikshit2021space, dikshit2020microbially, kumar2020bacterial}. All of the reviewed research show a compressive strength range from about 1 MPa - 200 MPa, with the majority falling under 100 MPa. The values found in this work are in agreement to the values found in other research. The small variance among the research out there is a result of differences in the manufacturing method.

\subsection{Shrinkage Behavior}

\begin{figure}[ht]
	\centering
		\includegraphics[scale=0.72]{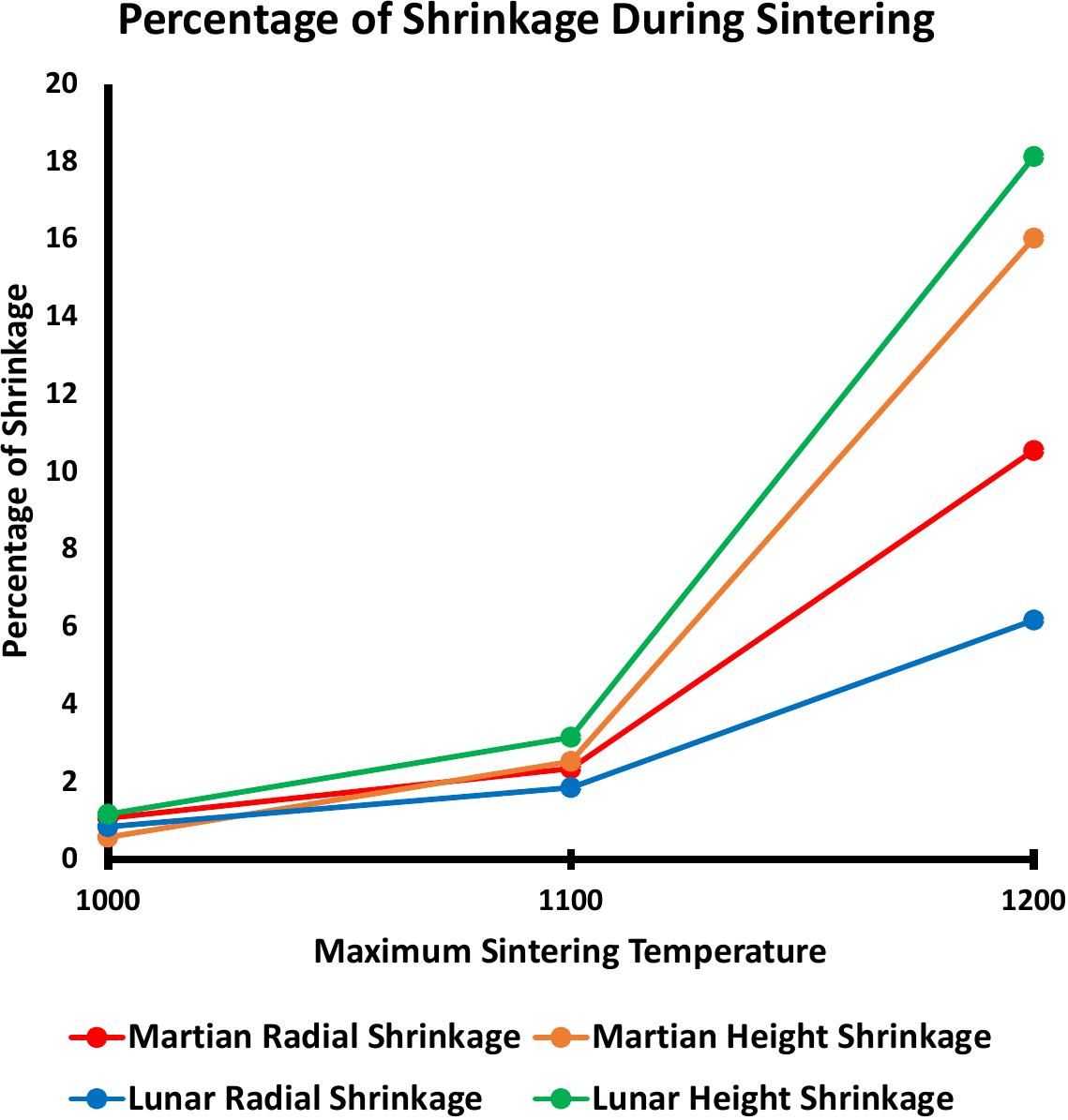}
	\caption{Plot of the percentage of shrinkage in both the radial and height direction that occurred during the sintering phases at 1000$^{\circ}$C, 1100$^{\circ}$C, and 1200$^{\circ}$C for both Martian and Lunar samples. The percentage of shrinkage indicates the amount of shrinkage that occurred from the green state to the sintered state.}
	\label{shrink}
\end{figure}

The shrinkage was calculated for all of the samples before and after sintering. This measurement was completed with digital calipers. There was negligible shrinkage for the samples sintered at 1000$^{\circ}$C and 1100$^{\circ}$C (about 1\% - 2\% in both directions). There was however a significant amount of shrinkage in both the radial and vertical directions for the samples sintered at 1200$^{\circ}$C. These values are reported in Table \ref{tab:MTS}. The values are also plotted in Figure \ref{shrink}. The Lunar samples sintered at 1200$^{\circ}$C  experienced more shrinkage in the vertical direction than the Martian samples sintered at the same conditions. The Lunar samples sintered at 1200$^{\circ}$C also experienced less radial shrinkage than the Martian samples sintered at the same conditions. All of the samples were sintered standing straight up, as seen in Figure \ref{Furn}. 

The dimensional shrinkage reported in this work has a statistical weight of 5 for both the Lunar and Martian samples at each sintering condition, because 5 samples were manufactured at each condition. If this material is to be additively manufactured via BJT the shrinkage behavior has to be accurately predicted prior to sintering and scaled up accordingly. This is already a common practice in BJT additive manufacturing for both metallic and ceramic parts \cite{wang2017investigation, bai2017effect}. Shrinkage in each direction must be accurately predicted. It is clear from this work that inequalities between vertical and horizontal shrinkage are not negligible for both the Martian and Lunar samples.  

\subsection{Mineral Composition}\label{Composition}

\begin{figure}[ht]
	\centering
		\includegraphics[scale=0.60]{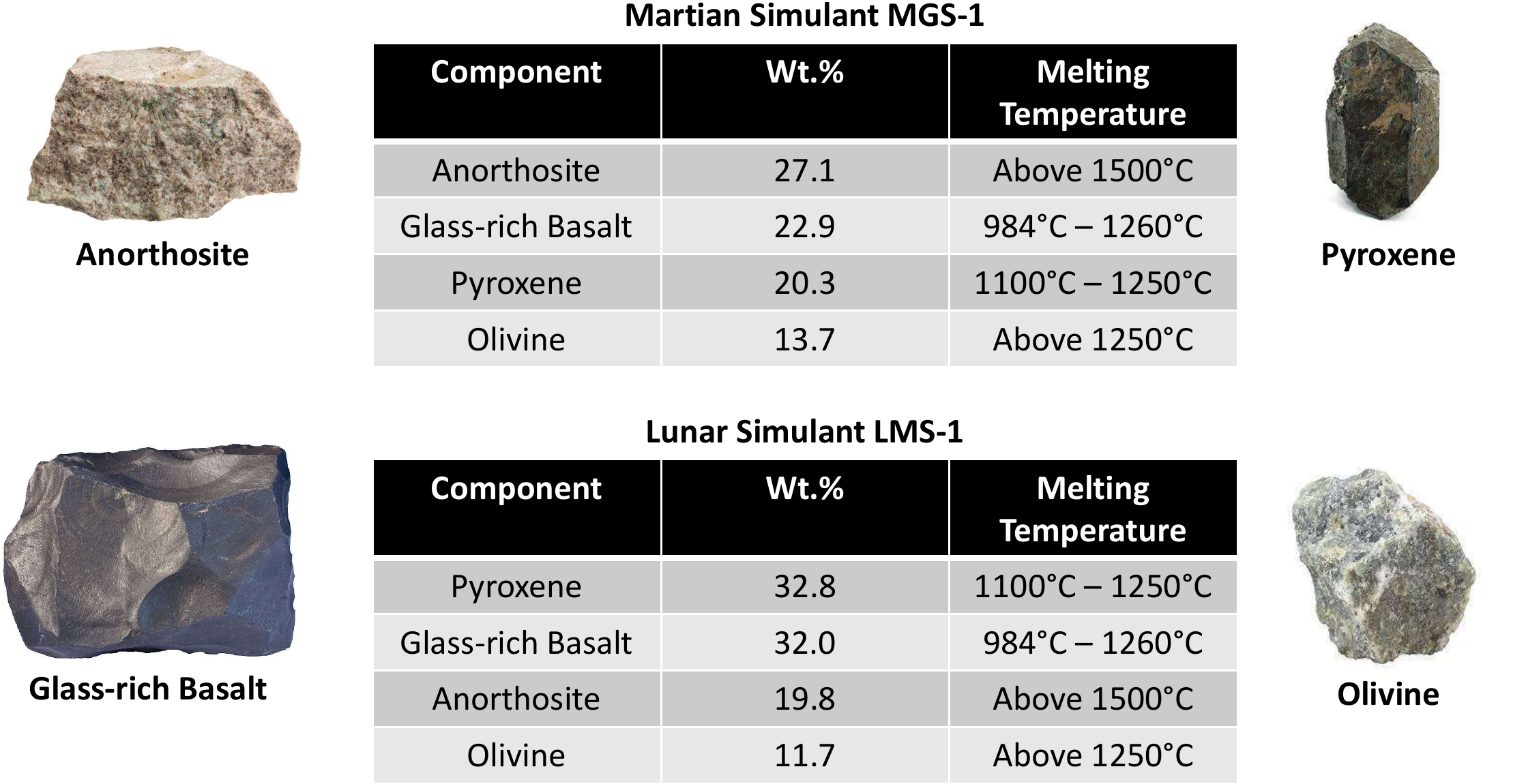}
	\caption{The 4 most common components are listed for the Lunar Simulant (LMS-1) and the Martian Simulant (MGS-1). Also listed is the weight percentage the component makes up of the respective simulant and the melting temperature range for each component. An example picture is also given for each of the components.}
	\label{Comp}
\end{figure}

The mineral composition that is found in the regolith is shown in Figure \ref{Comp} \cite{cannon2019mars, landsman2020simulated}. This figure only displays the top 4 components of the simulants (which accounts for about 80\% - 90\% of the total composition simulants). A full chemical and mineral composition of the simulants is available on the Exolith Labs simulant website \cite{Exolith}. The melting points for each of the components are displayed in the Figure \ref{Comp}. For both the Martian and the Lunar simulant, the glass-rich basalt and the pyroxene are the only components that are reaching to the melting temperatures at the sintering temperatures which were utilized for this work. Figure \ref{shrink} shows a substantial increase in shrinkage that occurs somewhere in between 1100$^{\circ}$C and 1200$^{\circ}$C.

The Lunar samples underwent more vertical shrinkage during sintering than the Martian samples. The Lunar samples contain about 10\% more glass-rich basalt than the Martian samples. This data demonstrates that the percentage of glass-rich basalt is a factor in the sintering process. The glass rich basalt does have the lowest melting point of all of the major components for both Martian and Lunar regolith.

\section{Conclusion}

This study has shown the both Martian and Lunar regolith have the potential to be implemented in an additive manufacturing method referred to as binder jetting. The binder material used in this work was a 6.25\% salt water solution, and it held the material together adequately. The simulants did not undergo any preprocessing procedures for this study. After sintering it was found that the compressive strength and compressive modulus of the Martian samples was slightly higher than the Lunar samples. The compressive strength and compressive modulus of both the Lunar and Martian samples was found to be in the range of brick, which indicates that the material is capability for the construction of load bearing structures. The material failure mode and stress strain behavior showed that the material is brittle and ceramic, for both the Lunar and Martian samples. The DIC data also showed sudden catastrophic failure in the form of vertical crack formation and propagation. This method of additive manufacturing could be implemented in-situ in extraterrestrial environments. The final printed and sintered components would be well suited to laying foundation, erecting structures, and fabricating miscellaneous components. 

It was also determined that the most optimal sintering temperature is somewhere between 1100$^{\circ}$C and 1200$^{\circ}$C. The shrinkage was around 15\% - 20\% in the vertical directions and 5\% - 10\% in the radial direction when sintered at 1200$^{\circ}$C. Most of the shrinkage during sintering occurred between the temperatures of 1100$^{\circ}$C and 1200$^{\circ}$C. The Lunar samples did undergo more shrinkage and deformation during sintering, this indicates a lower sintering temperature will be required for the Lunar samples for optimal properties. The Martian material had a slightly higher compression modulus and compressive strength than the Lunar material under the given manufacturing conditions. In the future, a more robust study of the sintering parameters should be completed, this includes both temperature and duration.

The microstructure of the samples was also analyzed via microscope at each of the manufacturing stages. After it was determined that water alone would not work as a binder material for either Martian or Lunar regolith, a 6.25\% salt-water solution was used. The bond that occurred between the particles in both powders was seen to be driven primarily by the smaller particles (less than 50 microns). This mixture of binder and small particles is what holds the larger particles together. It would be beneficial to explore the effect of milling the regoliths (both Martian and Lunar) prior to manufacturing to decrease the average size of the particles that make up the powder. It is also seen that the samples that were sintered at 1200$^{\circ}$C are very glassy in appearance. This is due to the presence of glass rich basalt in the regoliths. The glass-rich basalt is in the melting range during sintering at 1200$^{\circ}$C. It was shown in this work, through microstructural images and mechanical testing that the percentage of glass-rich basalt that comprises the regolith effects the sintering behavior. The higher quantity of glass rich basalt in the Lunar regolith led to an increase in deformation during sintering that occurred for the Lunar samples.

The cost of transportation of earth based materials is astronomically higher than simply utilizing materials that are readily available in these environments. The need for adaptive manufacturing in these challenging environments is critical to design and construct habitats capable of housing equipment and possibly people. This need can be met through the implementation of additive manufacturing. The development of additive manufacturing methods which utilize materials found in such environments would greatly assist in the exploration and establishment of settlements in these unforgiving environments. Additive manufacturing provides much needed versatility in an extraterrestrial setting. The ability to manufacture components rapidly to solve challenges that may arise on the Moon or on Mars is vital to ensuring success of such extreme endeavors.

\section{Acknowledgements}

We would like to thank Dr. Danial Britt and Exolith Labs at the University of Central Florida for assistance with the procurement of both Martian and Lunar regolith. %They were extremely helpful during our research and provided us with the regolith that was used for our experiments.

%% If you have bibdatabase file and want bibtex to generate the
%% bibitems, please use
%%
 \bibliographystyle{elsarticle-num} 
 \bibliography{cas-refs}

%% else use the following coding to input the bibitems directly in the
%% TeX file.

% \begin{thebibliography}{00}

% %% \bibitem{label}
% %% Text of bibliographic item

% \bibitem{}

% \end{thebibliography}
\end{document}